\documentclass[12pt]{article}
\usepackage{amsmath,amsfonts,amssymb,bbm}
\usepackage{tensor}
\usepackage[T1]{fontenc}
\usepackage[utf8]{inputenc}
\usepackage{lmodern}
\usepackage{authblk}
\usepackage[disable]{todonotes}
\usepackage{color}
\usepackage{multicol}
\usepackage{cite}
\usepackage[margin=3cm]{geometry}
\usepackage{hyperref}

\newcommand{\bz}{\bar{z}}

\newcommand{\p}{\partial}

\newcommand{\be}{\begin{equation}}
\newcommand{\ee}{\end{equation}}
\def\bea{\begin{eqnarray}}
\def\eea{\end{eqnarray}}
\def\nn{\nonumber}
\def\bm{\bar{m}}
\newcommand{\half}{\frac12}
\def\n{\nabla}
\def\bP{\bar P}
\def\bQ{\bar Q}
\def\bp{\bar\p}
\def\bz{\bar{z}}
\def\bY{\bar{Y}}
\def\bA{\bar{A}}
\def\bB{\bar{B}}
\def\bL{\bar{L}}

\allowdisplaybreaks[1]

\newcommand*\xbar[1]{%
  \hbox{%
    \vbox{%
      \hrule height 0.5pt 
      \kern0.3ex
      \hbox{%
        \kern-0.0em
        \ensuremath{#1}%
        \kern-0.0em
      }%
    }%
  }%
}
\def\pu{{}}
\def\pj{{}}

\newcommand\email[1]{\thanks{\href{mailto:#1}{\nolinkurl{#1}}}}

\author[a,b]{Pujian Mao}

\affil[a]{\,Center for Joint Quantum Studies and Department of Physics, School of Science, Tianjin University, 135 Yaguan Road, Tianjin 300350, P. R. China}

\affil[b]{\,Institute of High Energy Physics and Theoretical Physics Center for Science Facilities, Chinese Academy of Sciences, 19B Yuquan Road, Beijing 100049, P. R. China}

\title{\bf Asymptotics with a cosmological constant: The solution space\\}
\date{}

\begin{document}

\maketitle
\thispagestyle{empty}

\begin{abstract}
In this work, the solution space in the Newman-Penrose formalism with a cosmological constant is derived. The residual gauge transformation preserving the solution space is also worked out. By turning off the cosmological constant, the solution space has a well-defined flat limit. The asymptotic symmetry group of the resulting solution space consists of Diff($S^2$) transformations and supertranslations.
\end{abstract}


\section{Introduction}

Up until the 1960s, the existence of gravitational waves was still debatable. It was not clear if the radiation was just an artifact of linearization. To understand the gravitational radiation in full Einstein theory, Bondi, van der Burg, and Metzner established an elegant framework of expansions for axi-symmetric isolated systems \cite{Bondi:1962px}. In a suitable coordinates system, the metric fields are expanded in inverse powers of a radius coordinate\footnote{The problem of convergence of this expansion was fixed by Friedrich 20 years later \cite{Friedrich:1982}.}. Then, one can solve the equations of motion order by order with respect to proper boundary conditions that are asymptotically approaching flatness. In this framework, the radiation is characterized by a single function of two variables, the so-called \textit{news} functions; the mass of a system always decreases when there is news. In the same year, Sachs extended this framework to asymptotically flat space-times \cite{Sachs:1962wk}. Meanwhile, Newman and Penrose developed a new method to understand gravitational radiation by means of a tetrad or spinor formalism \cite{Newman:1961qr}. The peeling property of the Weyl tensors and the geometric meaning of certain gauge choice become transparent in the Newman-Penrose (NP) formalism.

The success in the understanding of gravitational radiation in full Einstein theory only refers to the case without a cosmological constant, $\Lambda=0$. In the presence of a cosmological constant, \pu{the radiation field is essentially origin dependent \cite{Penrose:1965am}. In particular, when $\Lambda$ is positive, the radiation vanishes along directions opposite to principal null directions \cite{Krtous:2003rw}. Hence, the absence of certain radiation field can not be used to distinguish nonradiative sources. When $\Lambda$ is negative,} an extra requirement, the reflective boundary condition, is imposed to make the evolution well defined \cite{Hawking:1983mx}. Consequently, there is no analogue of the Bondi news functions \cite{Ashtekar:1984zz}. Gravitational radiation seems to be turned off by the appearance of a \pu{negative} cosmological constant. However, surprising results were obtained in the case with a positive cosmological constant recently \cite{Ashtekar:2014zfa,Ashtekar:2015lla,Ashtekar:2015ooa,Ashtekar:2015lxa}. A suitable framework which allows one to apply the late-time
and post-Newtonian approximations and finally to express the leading terms of solutions in terms of the quadrupole moments of sources is proposed (see also \cite{He:2015wfa,He:2017hmj,Szabados:2015wqa,Zhang:2015iua,Saw:2016isu,Date:2015kma,Bishop:2015kay,Xie:2017uqa,Chrusciel:2016oux,He:2016vji,Kesavan:2016pvc,Saw:2016pez,Date:2016uzr,Saw:2017zks,Bonga:2017dpl,Ashtekar:2017dlf,Saw:2017hsf,Bonga:2017dlx,Borghini:2017xew,Borghini:2017eqa,Saw:2017amv,He:2018ikd,Wu:2018cna,Szabados:2018erf,Poole:2018koa} for recent relevant developments).

According to the experience of the case $\Lambda=0$, a fully controlled solution space is the most important ingredient in the understanding of radiation and gravitationally conserved quantities, in both the metric formalism and NP formalism. \pu{Several works have been carried out in this direction. The solution space of asymptotically dS space-times was recently derived in \cite{Ashtekar:2014zfa} using geometric tools \`{a} la Penrose \cite{Penrose:1962ij,Penrose:1965am} with special emphasis on the consequence of a conformally flat boundary 2 metric. In the metric formalism, the axi-symmetric solution space in Bondi gauge was obtained with a cosmological constant in \cite{He:2015wfa,Poole:2018koa}. The solution space of the NP formalism with a cosmological constant was derived in \cite{Saw:2016isu,Saw:2017hsf} with a special choice of the foliation of space-times geometry which is a family of null hyper-surfaces given by constant values of the $u$ coordinate.} In this paper, we work out the \pu{most general} solution space of the NP formalism with a cosmological constant \pu{by removing the constant foliation condition in \cite{Saw:2016isu,Saw:2017hsf}}. \pu{The main reason to do so is to include the AdS Robinson-Trautman solution \cite{Bakas:2014kfa} which is an exact solution with spherical gravitational waves but does not satisfy the constant foliation condition in its simplest form.} In contrast with the case $\Lambda=0$, the news functions are completely determined by the data on the boundary 2 metric. Two functions from the boundary 2 metric are left undefined. They represent the radiation in this system. The residual gauge transformation that preserves the form of the solution space is also derived. The asymptotic symmetry group is the full diffeomorphism group of the boundary 2 surface. There is no analogue of supertranslation in the case with a cosmological constant \cite{Ashtekar:1984zz,Ashtekar:2014zfa}. We show in details the consequence of taking the flat limit, $\Lambda=0$. The resulting solution space includes the Newman-Unit (NU) solution space \cite{Newman:1962cia} as a particular case, \textit{i.e.} the boundary 2 metric to be conformally flat. Its residual gauge symmetry consists of Diff($S^2$) transformations and supertranslations.

The rest of this paper is organized as follows. In section 2, we will review the Newman-Penrose formalism. Section 3 will present the main result of this paper, the solution space of NP formalism with a cosmological constant. In section 4, we compute the residual transformation preserving the solution space. Section 5 will show the details of the flat limit. The last section is devoted to several open issues, some of which we hope to study elsewhere. There are also three Appendixes in which useful information for the main text is presented.

\section{Review of Newman-Penrose formalism}

Newman and Penrose \cite{Newman:1961qr} established a special tetrad formalism with four null basis vectors $e_1=l,\;e_2=n,\;e_3=m,\;e_4=\bar{m}$. The basis vectors $l$ and $n$ are real, while $m$ and $\bm$ are complex conjugates of each other. The null basis vectors have the orthogonality conditions
\be
l\cdot m=l\cdot\bm=n\cdot m=n\cdot\bm=0
\ee
and are normalized as
\be
l\cdot n=1,\;\;\;\;m\cdot\bm=-1.
\ee
The space-times metric is obtained from
\be\label{metric}
g_{\mu\nu}=n_\mu l_\nu + l_\mu n_\nu - m_\mu {\bm}_\nu - m_\nu \bm_\mu.
\ee
The connection coefficients are now called spin coefficients, labeled by several Greek symbols (we will follow the convention of \cite{Chandrasekhar}):
\begin{align}
&\kappa=\Gamma_{311}=l^\nu m^\mu\nabla_\nu l_\mu,\;\;\pi=-\Gamma_{421}=-l^\nu \bar{m}^\mu\nabla_\nu n_\mu,\nn\\
&\epsilon=\half(\Gamma_{211}-\Gamma_{431})=\half(l^\nu n^\mu\nabla_\nu l_\mu - l^\nu \bar{m}^\mu\nabla_\nu m_\mu),\nn\\
&\nn\\
&\tau=\Gamma_{312}=n^\nu m^\mu\nabla_\nu l_\mu,\;\;\nu=-\Gamma_{422}=-n^\nu \bar{m}^\mu\nabla_\nu n_\mu,\nn\\
&\gamma=\half(\Gamma_{212}-\Gamma_{432})=\half(n^\nu n^\mu\nabla_\nu l_\mu - n^\nu \bar{m}^\mu\nabla_\nu m_\mu),\nn\\
&\nn\\
&\sigma=\Gamma_{313}=m^\nu m^\mu\nabla_\nu l_\mu,\;\;\mu=-\Gamma_{423}=-m^\nu \bar{m}^\mu\nabla_\nu n_\mu,\label{coefficient}\\
&\beta=\half(\Gamma_{213}-\Gamma_{433})=\half(m^\nu n^\mu\nabla_\nu l_\mu - m^\nu \bar{m}^\mu\nabla_\nu m_\mu),\nn\\
&\nn\\
&\rho=\Gamma_{314}=\bar{m}^\nu m^\mu\nabla_\nu l_\mu,\;\;\lambda=-\Gamma_{424}=-\bar{m}^\nu \bar{m}^\mu\nabla_\nu n_\mu,\nn\\
&\alpha=\half(\Gamma_{214}-\Gamma_{434})=\half(\bar{m}^\nu n^\mu\nabla_\nu l_\mu - \bar{m}^\nu \bar{m}^\mu\nabla_\nu m_\mu).\nn
\end{align}
Ten independent components of the Weyl tensors are represented by five complex scalars,
\begin{align}
\Psi_0=-C_{1313},\;\;\Psi_1=-C_{1213},\;\;\Psi_2=-C_{1342},\;\;\Psi_3=-C_{1242},\;\;\Psi_4=-C_{2424}.\nn
\end{align}
Ricci tensors are defined in terms of four real and three complex scalars,
\begin{align}
&\Phi_{00}=-\half R_{11},\;\;\Phi_{22}=-\half R_{22},\;\;\Phi_{02}=-\half R_{33},\;\;\Phi_{20}=-\half R_{44},\nn\\
&\Phi_{11}=-\dfrac{1}{4}(R_{12}+R_{34}),\;\;\Phi_{01}=-\half R_{13},\;\;,\Phi_{12}=-\half R_{23},\nn\\
&\Lambda=\dfrac{1}{24}R=\dfrac{1}{12} (R_{12}-R_{34}),\;\;\Phi_{10}=-\half R_{14},\;\;\Phi_{21}=-\half R_{24},\nn
\end{align}
where $\Lambda$ is the cosmological constant.

In the NP formalism, by local Lorentz transformations, it is always possible to impose
\begin{align}
\pi=\kappa=\epsilon=0,\,\,\;\;\rho=\bar\rho,\;\;\,\,\tau=\bar\alpha+\beta.
\end{align}
According to \eqref{1} in Appendix \ref{relation}, such a gauge choice means that $l$ is tangent to a null geodesic with an affine parameter. Moreover, the congruence of the null geodesic is hypersurface orthogonal; namely, $l$ is proportional to the gradient of a scalar field. It is of convenience to choose this scalar field as coordinate $u=x^1$ and take the affine parameter as coordinate $r=x^2$. Then, to satisfy the orthogonality conditions and normalization conditions, the tetrad and the co-tetrad must have the forms
\begin{align}
&n=\frac{\p}{\p u} + U \frac{\p}{\p r} + X^A \frac{\p}{\p x^A},\nn\\
&\label{gauge}l=\frac{\p}{\p r},\;\;\;\;\;\;m=\omega\frac{\p}{\p r} + L^A \frac{\p}{\p x^A},\\
&n=\left[-U-X^A(\xbar\omega L_A+\omega \bar L_A)\right]du + dr + (\omega\bar L_A+\xbar\omega L_A)dx^A,\nn\\
&l=du,\;\;\;\;\;\;\;m=-X^AL_A du + L_A dx^A,\nn
\end{align}
where $L_AL^A=0,\;L_A\bar L^A=-1$. Considered as directional derivatives, the basis vectors are designated by special symbols:
\begin{align}
D=l^\mu\p_\mu,\;\;\;\;\Delta=n^\mu\p_\mu,\;\;\;\;\delta=m^\mu\p_\mu.
\end{align}
The full vacuum Newman-Penrose equations with a cosmological constant are listed in Appendix \ref{NP}.

\section{Solution space}
\label{solution}
The main condition that all components of the Weyl tensor ($\Psi$s) approach zero at infinity is $\Psi_0=\frac{\Psi_0^0}{r^5} + o(r^{-5})$. However, a slightly stronger condition that $\Psi_0=\frac{\Psi_0^0}{r^5} + \frac{\Psi_0^1}{r^6} + O(r^{-7})$ is usually adopted \cite{Newman:1962cia} to apply a $\frac1r$ series expansion. The asymptotically flat solution space of NP equations was derived by Newman and Unti \cite{Newman:1962cia} more than half century ago. The appearance of the cosmological constant has a significant consequence for the solution space. In \cite{Saw:2016isu}, the NP equations are solved in $(\theta,\phi)$ coordinates with \pu{a special choice of the foliation of space-times geometry and a third class of tetrad rotation to set the order 1 piece of the spin coefficient $\gamma$ to be zero ($\gamma^0=0$).} \pj{Indeed, it is always possible to set $\gamma^0=0$ in the $\Lambda=0$ case by residual gauge transformations, see, for instance, \cite{vonderGonna:1997sh,Adamo:2009vu} for the precise coordinates transformations. However, as shown in the next section, such a type of gauge transformations will involve logarithm terms in nonzero $\Lambda$ case. Since only $\frac1r$ expansions are assumed in \cite{Saw:2016isu}, the $\gamma^0=0$ condition will eliminate the possible situation that represents gravitational radiation, for instance the (A)dS Robinson-Trautman solutions \cite{Bakas:2014kfa} (see also its application in AdS/CFT therein). We will further comment on this point in section \ref{flat}.} \pu{In addition, stereographic coordinates $(z,\bz)$ are more convenient for certain issues, for instance, the recent attempts at flat-space holography \cite{Barnich:2009se,Barnich:2010eb} and the triangle equivalence \cite{Strominger:2017zoo}.} \pj{In this section, we will work out the most general solution space of NP equations with a cosmological constant in stereographic coordinates.}

\pu{The derivation of the solution space is explained in Appendix \ref{process} in details. We summarize as follows}: the process of solving the radial equations \eqref{R1}-\eqref{R17} is exactly the same as in \cite{Newman:1962cia}. The cosmological constant will just modify the asymptotic behaviors of the spin coefficients as
\begin{align}
\lambda=\Lambda\xbar\sigma^0+O(r^{-1}),\;\;\;\;\mu=\Lambda r+O(r^{-1}),\;\;\;\;\gamma=-\Lambda r+O(r^0),
\end{align}
which agree with the results in \cite{Saw:2016isu}\footnote{\pu{To compare with the result in \cite{Saw:2016isu}, one needs to do the following replacement: $\pi\rightarrow-\tau',\;\epsilon\rightarrow-\gamma',\;\nu\rightarrow-\kappa',\;\mu\rightarrow-\rho',\;\beta\rightarrow-\alpha',\;\lambda\rightarrow-\sigma',
\;\Lambda\rightarrow\frac{\Lambda}{6},\;\Psi$s$\rightarrow-\Psi$s.}}. However the constraints from non-radial equations on the integration constants of the radial equations are much stronger than in the $\Lambda=0$ case. In particular, $\sigma^0$ and $\gamma^0$ are completely determined by \eqref{H8} as
\be\label{sigma}
\Lambda\sigma^0=\frac{Q\p_u P -P\p_u Q}{2(P\bP - Q\bQ)},\;\;\;\;\gamma^0=\frac{Q\p_u \bQ -P\p_u \bP}{2(P\bP - Q\bQ)},
\ee
where $Q(u,z,\bz)$ and $P(u,z,\bz)$ are the integration constants in $L^z$ and $L^{\bz}$. $P$ and $Q$ are the only variables of which the time evolutions are not determined. They represent gravitational radiation in this system. This is in good consistency with that of \cite{Ashtekar:2014zfa,Saw:2016isu}.

Before showing the full solutions, it is very useful to introduce the ``$\eth$'' operator that is defined as
\begin{equation}\begin{split}
&\eth \eta^{(s)}=L^A_0 \frac{\p}{\p x^A} \eta^{(s)} + 2 s\xbar\alpha^0 \eta^{(s)},\\
&\xbar\eth \eta^{(s)}=\xbar L^A_0 \frac{\p}{\p x^A} \eta^{(s)} -2 s \alpha^0 \eta^{(s)},
\end{split}\end{equation}
where $L^A_0$ is the integration constant in $L^A$ and $s$ is the spin weight of the field $\eta$. The spin weights of relevant fields are listed below in Table \ref{t1}.
\begin{table}[h]
\caption{Spin weights}\label{t1}
\begin{center}\begin{tabular}{|c|c|c|c|c|c|c|c|c|c|c|c|c|c|c|c|c|c}
\hline
& $\eth$ & $\p_u$ & $\gamma^0$ & $\nu^0$ & $\mu^0$ & $\sigma^0$ & $\lambda^0$  & $\Psi^0_4$ &  $\Psi^0_3$ & $\Psi^0_2$ & $\Psi^0_1$ & $\Psi_0^0$  & $P$ &$Q$ \\
\hline
s & $1$& $0$& $0$& $-1$& $0$& $2$& $-2$  &  $-2 $&  $-1$ & $0$ & $1$ & $2$  & $1$  &  $1$  \\
\hline
\end{tabular}\end{center}\end{table}

The full solutions of NP equations with a cosmological constant in asymptotic expansions are given by:
\begin{align}
&\Psi_0=\dfrac{\Psi_0^0}{r^5}+ \dfrac{\Psi_0^1}{r^6} +O(r^{-7}),\label{psi0}\\
&\Psi_1=\dfrac{\Psi_1^0}{r^4}-\dfrac{\xbar \eth \Psi_0^0}{r^5}+O(r^{-6}),\label{psi1}\\
&\Psi_2=\dfrac{\Psi_2^0}{r^3}+\dfrac{\Lambda \xbar\sigma^0\Psi_0^0-\xbar \eth \Psi_1^0}{r^4}+O(r^{-5}),\label{psi2}\\
&\Psi_3=\dfrac{\Psi_3^0}{r^2}+\dfrac{2\Lambda \xbar\sigma^0\Psi_1^0-\xbar \eth\Psi_2^0}{r^3}+O(r^{-4}),\label{psi3}\\
&\Psi_4=\dfrac{\Psi_4^0}{r}+\dfrac{3\Lambda \xbar\sigma^0\Psi_2^0-\xbar \eth \Psi_3^0}{r^2}+O(r^{-3}),\label{psi4}\\
&\nn\\
&\rho=-\frac{1}{r}-\frac{\sigma^0\xbar\sigma^0}{r^3}+O(r^{-5}),\label{rho}\\
&\sigma=\frac{\sigma^0}{r^2} + \frac{{\sigma^0}\sigma^0\xbar\sigma^0 - \frac12\Psi_0^0}{r^4} -\frac{\Psi_0^1}{3r^5} +O(r^{-6}),\label{sigma}\\
&\label{alpha}\alpha=\frac{\alpha^0}{r}+\frac{\xbar\sigma^0\xbar\alpha^0}{r^2}+\frac{\sigma^0\xbar\sigma^0\alpha^0}{r^3}+O(r^{-4}),\;\;\;\;\;\;\;\;\tau=-\frac{\Psi^0_1}{2r^3}+O(r^{-4}),\\
&\label{beta}\beta=-\frac{\xbar\alpha^0}{r}-\frac{\sigma^0\alpha^0}{r^2}-\frac{\sigma^0\xbar\sigma^0\xbar\alpha^0+\half \Psi^0_1}{r^3}+O(r^{-4}),\\
&\label{mu}\mu=\Lambda r + \frac{\mu^0}{r} - \frac{\sigma^0\lambda^0+\Psi^0_2}{r^2} + \frac{\mu^0\sigma^0\xbar\sigma^0 + \frac12\xbar\eth\Psi^0_1 + \frac13\Lambda(\xbar\sigma^0\Psi_0^0 + \sigma^0\xbar\Psi_0^0)}{r^3} +O(r^{-4}),\\
&\label{lambda}\lambda=\Lambda \xbar\sigma^0 + \frac{\lambda^0}{r}-\frac{\xbar\sigma^0 \mu^0}{r^2} + \frac{\lambda^0\xbar\sigma^0 \sigma^0 + \frac12\xbar\sigma^0\Psi_2^0 + \frac16 \Lambda \xbar\Psi_0^1}{r^3} +O(r^{-4}),\\
&\label{gamma}\gamma=-\Lambda r + \gamma^0 - \frac{\Psi^0_2}{2r^2} + \frac{2\xbar\eth\Psi_1^0 + \alpha^0\Psi_1^0 - \xbar\alpha^0\xbar\Psi_1^0 - 2\Lambda\xbar\sigma^0\Psi_0^0}{6r^3}+O(r^{-4}),\\
&\label{nu}\nu=\nu^0+\frac{\frac12\Lambda\xbar\Psi_1^0-\Psi^0_3}{r}+\frac{3\xbar \eth \Psi^0_2-\Lambda \eth\xbar\Psi_0^0-5\Lambda\xbar\sigma^0\Psi_1^0}{6r^2}+O(r^{-3}),\\
&\nn\\
&\label{XA}X^z=\frac{\bP \Psi_1^0 + Q \xbar\Psi_1^0}{6r^3} + O(r^{-4}),\\
&\label{omega}\omega=\frac{\xbar \eth \sigma^0}{r}-\frac{\sigma^0\eth \xbar\sigma^0+\half \Psi^0_1}{r^2}+O(r^{-3}),\\
&\label{U}U=\Lambda r^2 - r(\gamma^0+\xbar\gamma^0) + U^0-\frac{\Psi^0_2 + \xbar \Psi^0_2}{2r}+ \frac{\xbar\eth\Psi_1^0 + \eth\xbar\Psi_1^0 - \Lambda\sigma\xbar\Psi_0^0 - \Lambda\xbar\sigma^0\Psi_0^0}{6r^2}+ O(r^{-3}),\\
&\label{Lz}L^{z}=\frac{Q}{r} - \frac{\bP\sigma^0}{r^2} +\frac{Q\sigma^0 \xbar\sigma^0}{r^3} +\frac{\bP(\Psi_0^0 - 6{\sigma^0}^2\xbar\sigma^0)}{6r^4} +O(r^{-5}),\\
&\label{Lzb}L^{\bz}=\frac{P}{r} - \frac{\bQ\sigma^0}{r^2} +\frac{P\sigma^0 \xbar\sigma^0}{r^3} + \frac{\bQ(\Psi_0^0 - 6{\sigma^0}^2\xbar\sigma^0)}{6r^4} +O(r^{-5}),\\
&\label{Lzd}L_z=\frac{-Pr}{P\bP-Q\bQ} + \frac{\bQ\sigma^0}{P\bP-Q\bQ} - \frac{1}{6r^2}\frac{\bQ\Psi_0^0}{P\bP-Q\bQ} + \frac{1}{12r^3}\frac{P\xbar\sigma^0\Psi_0^0-\bQ\Psi_0^1}{P\bP-Q\bQ}+ O(r^{-4}),\\
&\label{Lzbd}L_{\bz}=\frac{Qr}{P\bP-Q\bQ} - \frac{\bP\sigma^0}{P\bP-Q\bQ} + \frac{1}{6r^2}\frac{\bP\Psi_0^0}{P\bP-Q\bQ} - \frac{1}{12r^3}\frac{Q\xbar\sigma^0\Psi_0^0-\bP\Psi_0^1}{P\bP-Q\bQ}+ O(r^{-4}),
\end{align}
where
\begin{align}
&\label{alpha0}\alpha^0=\frac12 \bL_{0B}(L_0^A \p_A \bL_0^B - \bL_0^A \p_A L_0^B),\\
&\label{sigma0}\Lambda \sigma^0 =\frac12 L_{0A} \p_u L_0^A,\\
&\label{gamma0}\gamma^0=\frac12 L_{0A} \p_u \bL_0^A,\\
&\label{mu0}\mu^0=-\eth \alpha^0-\xbar\eth\xbar\alpha^0-2\Lambda\sigma^0\xbar\sigma^0,\\
&\label{lambda0}\lambda^0=\p_u\xbar\sigma^0+(3\gamma^0-\xbar\gamma^0)\xbar\sigma^0,\\
&\label{nu0}\nu^0=\xbar \eth (\gamma^0+\xbar\gamma^0)-2\Lambda\eth\xbar\sigma^0,\\
&\label{U0}U^0=\mu^0-\Lambda\sigma^0\xbar\sigma^0,\\
&\label{omega0}\omega^0=\xbar\eth\sigma^0,\\
&\label{Psi20}\Psi_2^0 - \xbar\Psi_2^0 = \xbar\lambda^0\xbar\sigma^0 - \lambda^0\sigma^0 + \xbar\eth^2\sigma^0-\eth^2\xbar\sigma^0,\\
&\label{Psi30}\Psi^0_3=\xbar \eth \mu^0 - \eth\lambda^0 + \Lambda \xbar\eth\xbar\sigma^0 - \Lambda \xbar \Psi_1^0,\\
&\label{Psi40}\Psi^0_4=\xbar\eth \nu^0 - \p_u\lambda^0 - 4\gamma^0\lambda^0 - 4 \Lambda \mu^0\xbar\sigma^0+ \Lambda^2 \xbar \Psi_0^0,
\end{align}
and the time evolutions of the Weyl tensors
\be\begin{split}\label{evolution}
&\p_u\Psi^0_0 + (\gamma^0 + 5 \xbar \gamma^0)\Psi^0_0=\eth\Psi^0_1 + 3\sigma^0\Psi_2^0 + \Lambda \Psi_0^1,\\
&\p_u\Psi^0_1 + 2 (\gamma^0 + 2 \xbar \gamma^0)\Psi^0_1=\eth\Psi^0_2 + 2\sigma^0\Psi_3^0 - \Lambda \xbar\eth\Psi_0^0 ,\\
&\p_u\Psi^0_2 + 3 (\gamma^0 + \xbar \gamma^0)\Psi^0_2=\eth\Psi^0_3  + \sigma^0\Psi_4^0 - \Lambda \xbar\eth\Psi_1^0 + \Lambda^2\xbar\sigma^0\Psi_0^0,\\
&\p_u\Psi^0_3 + 2 (2 \gamma^0 + \xbar \gamma^0)\Psi^0_3=\eth\Psi^0_4 - \Lambda \xbar\eth\Psi_2^0 + 2\Lambda^2\xbar\sigma^0\Psi_1^0,
\end{split}\ee
as well as the identities
\be\begin{split}\label{identities}
&\p_u \mu^0=\eth\xbar\eth(\gamma^0+\xbar\gamma^0)-2(\gamma^0+\xbar\gamma^0)\mu^0-\Lambda(\eth^2\xbar\sigma^0 + \xbar\eth^2\sigma^0) -2\Lambda(\xbar\lambda^0\xbar\sigma^0 + \lambda^0\sigma^0),\\
&\p_u\alpha^0=\Lambda\eth\xbar\sigma^0+2\Lambda\xbar\alpha^0\xbar\sigma^0-2\gamma^0\alpha^0-\xbar\eth\xbar\gamma^0.
\end{split}\ee
The commutator of the $\eth$ operator is\footnote{One should use the identity $\eth \bL_0^A=\xbar\eth L_0^A$.}
\be
[\xbar\eth,\eth]\eta^{(s)}=2s(\eth \alpha^0 + \xbar \eth \xbar\alpha^0)\eta^{(s)}.
\ee

\pu{The solution space in metric form can be easily obtained by inserting the tetrad solution \eqref{XA}-\eqref{Lzbd} into \eqref{metric}. The gauge choice in \eqref{gauge} is the so-called Newman-Unti gauge \cite{Newman:1962cia}. To compare to the results in Bondi gauge in \cite{He:2015wfa,Poole:2018koa}, one just needs to apply a transformation in the radial direction \cite{Barnich:2011ty}. A conformally flat boundary 2 metric explored in \cite{Ashtekar:2014zfa} corresponds to $\Psi_4^0=\Psi_3^0=$Im$(\Psi_2^0)=0$ in the NP formalism. Those conditions will eventually yield $Q=\p_u P=0$ in the above solution space. There is a very interesting exact solution with gravitational waves in truncated forms, \textit{i.e.} the (A)dS Robinson-Trautman solution \cite{Bakas:2014kfa} (see also earlier references therein). It is a generalization of the Robinson-Trautman waves \cite{Robinson:1960zzb} to the nonzero $\Lambda$ case. Adapted to our convention, the metric of the solution is}\footnote{We use $(+,-,-,-)$ signature. $\p$ and $\xbar\p$ denote $\p_z$ and $\p_{\bz}$, respectively. \pj{We follow the convention of \cite{Chandrasekhar} in which Einstein's equations with a cosmological constant in the metric formalism are $R_{\mu\nu}-\frac12 g_{\mu\nu} R + 6 \Lambda g_{\mu\nu} =0$. To compare with the metric in \cite{Bakas:2014kfa}, one needs to do the following replacement: $P\rightarrow e^{-\frac{\Phi}{2}},
\;\Lambda\rightarrow\frac{\Lambda}{6},\;\Psi_2^0\rightarrow-m$.}}
\pu{\be
ds^2=2\left(-\Lambda r^2-r\p_u \ln P + P^2\p \bp \ln P + \frac{\Psi_2^0}{r} \right)du^2+2dudr-2\frac{r^2}{P^2}dzd\bz,
\ee}
where $P$ is a real function of $(u,z,\bz)$ and $\Psi_2^0$ is a real constant. The time evolution equation of the conformal factor is
\be\label{RTequation}
3\Psi^0_2\p_u P + P^3\bp^2 P \p^2 P - P^4\p^2\bp^2 P=0.
\ee
In the NP formalism, the solution is given by
\begin{align}\label{RT}
&\Psi_0=\Psi_1=\sigma=\lambda=\tau=X^A=\omega=Q=0,\nn\\
&\Psi_2=\frac{\Psi_2^0}{r^3},\;\;\Psi_2^0 \text{ is a real constant},\nn\\
&\Psi_3=\frac{P\p \mu^0}{r^2},\;\;\;\;\Psi_4=\frac{-\p(P^2\p\p_u\ln P)}{r}-\frac{P^2\p\bp\mu^0}{r^2},\;\;\;\;\mu^0=-P^2 \p\xbar\p \ln P\nn\\
&\rho=-\frac{1}{r},\;\;\;\;\alpha=\frac{\p P}{2r},\;\;\;\;\beta=-\frac{\bp P}{2r},\;\;\mu=\Lambda r+\frac{\mu^0}{r} - \frac{\Psi^0_2}{r^2},\\
&\gamma=-\Lambda r-\half \p_u \ln P-\frac{\Psi^0_2}{2r^2},\;\;\nu=-P\p\p_u\ln P-\frac{P\p \mu^0}{r},\nn\\
&U=\Lambda r^2 + \p_u \ln P r + \mu^0-\frac{\Psi^0_2}{r},\;\;L^z=0,\;\;L^{\bz}=\frac{P}{r},\nn
\end{align}

\section{Residual gauge transformation}
\pj{In this section, we will work out the residual gauge transformation preserving the solution space derived in the previous section. We follow closely the process of \cite{Barnich:2016lyg}, in which the residual gauge transformation preserving the forms of the NU solution space \cite{Newman:1962cia} was studied in detail (see also earlier references therein).} A gauge transformation of the first order formalism of Einstein gravity is a combination of a change of coordinates and a local Lorentz transformation. In the NP formalism, the local Lorentz transformation is described in the standard three classes of rotation \cite{Chandrasekhar}. A combined rotation $II\circ I \circ III$ of the tetrad basis is given by\footnote{To connect with the notation in \cite{Chandrasekhar}, one just needs to set $A\leftrightarrow a,\;B\leftrightarrow b,\;e^{-E_R}\leftrightarrow A,\;E_I\leftrightarrow\theta$.}
\begin{align}\label{tt}
&\tilde{l}=(1+\bA B)(1 + A \bB)e^{E_R}l + B \bB e^{-E_R} n + \bB(1+\bA B)e^{iE_I}m + B(1+A\bB)e^{-iE_I}\bm,\nn\\
&\tilde{n}=A\bA e^{E_R}l + e^{-E_R}n + \bA e^{iE_I}m + A e^{-iE_I}\bm,\\
&\tilde{m}=A(1+\bA B)e^{E_R}l + Be^{-E_R}n + (1+\bA B)e^{iE_I} m + ABe^{-iE_I}\bm.\nn
\end{align}
The change of coordinates is in the form
\be
u=u(u',r',z',\bz'),\;\;r=r(u',r',z',\bz'),\;\;x^A=x^A(u',r',z',\bz').
\ee

The gauge condition $l'=\frac{\p}{\p r'}$ implies
\be
\frac{\p x^\mu}{\p r'}=\tilde l^\mu.
\ee
Hence,
\begin{align}
&\frac{\p u}{\p r'}=B \bB e^{-E_R},\nn\\
&\frac{\p z}{\p r'}=X^z B \bB e^{-E_R}+L^z \bB(1+\bA B)e^{iE_I} + \bL^z B(1+A\bB)e^{-iE_I},\\
&\frac{\p r}{\p r'}=(1+\bA B)(1 + A \bB)e^{E_R}+U B \bB e^{-E_R} + \omega \bB(1+\bA B)e^{iE_I} + \xbar\omega B(1+A \bB)e^{-iE_I},\nn
\end{align}
which is (6.41) in \cite{Barnich:2016lyg}. This will fix the unprimed coordinates up to 4 integration constants of $r'$. To implement the gauge condition $\kappa=\pi=\epsilon=0$, one has to go back to the original definition in \eqref{coefficient}. The transformed spin coefficients are
\be
\kappa'=-\tilde{l}^\nu \tilde{l}^\mu \nabla_\nu \tilde{m}_\mu,\;\;\xbar\pi'=-\tilde{l}^\nu \tilde{m}^\mu\nabla_\nu \tilde{n}_\mu,\;\;\epsilon'=\tilde{l}^\nu \tilde{n}^\mu\nabla_\nu \tilde{l}_\mu.
\ee
\pu{Inserting \eqref{tt} into the transformed spin coefficients and} applying the relation in Appendix \ref{relation}, the gauge conditions $\kappa'=\pi'=\epsilon'=0$ eventually lead to
\begin{align}
&\tilde{D}B= e^{E} \left[B \bB e^{-E_R} \tau + \bB(1+\bA B)e^{iE_I}\sigma + B(1+A\bB)e^{-iE_I}\rho\right],\nn\\
&\tilde{D}\bA=\bA^2e^{E} \left[B \bB e^{-E_R} \tau + \bB(1+\bA B)e^{iE_I}\sigma + B(1+A\bB)e^{-iE_I}\rho\right]\nn\\
&\hspace{2cm}- e^{-E} \left[B \bB e^{-E_R} \nu + \bB(1+\bA B)e^{iE_I}\mu + B(1+A\bB)e^{-iE_I}\lambda\right],\label{Lorentz}\\
&\tilde{D}E=-2\bA e^{E} \left[B \bB e^{-E_R} \tau + \bB(1+\bA B)e^{iE_I}\sigma + B(1+A\bB)e^{-iE_I}\rho\right]\nn\\
&\hspace{2cm}- 2 \left[B \bB e^{-E_R} \gamma + \bB(1+\bA B)e^{iE_I}\beta + B(1+A\bB)e^{-iE_I}\alpha\right].\nn
\end{align}
where $\tilde{D}=\tilde{l}^\mu\p_\mu$. In the primed coordinates, $\tilde{D}'=\frac{\p}{\p r'}$ because of the gauge condition on $l$. Thus, \eqref{Lorentz} is precisely (6.44) in \cite{Barnich:2016lyg}. In the flat case, the following asymptotic behavior is assumed:
\be\begin{split}
&r=e^{E_{R0}}r'+O(1),\;\;\;\;u=O(1),\;\;\;\;x^A=O(1),\\
&A=O(1),\;\;\;\;E=O(1),\;\;\;\;B=O(r^{-1}).
\end{split}\ee
The three classes of rotation will be fixed up to 6 integration constants of $r'$. However, the appearance of the cosmological constant will involve logarithms of $r$ in the third equation of \eqref{Lorentz} because $\gamma=-\Lambda r+O(r^0)$. Then, logarithmic terms will show up in the transformed solution space which violates the assumption that the solution is given in $\frac1r$ expansion and should be ruled out\footnote{Expansion involving logarithms is called polyhomogeneous expansion \cite{Winicour,Chrusciel:1993hx,ValienteKroon:1998vn}. It is of interest to explore such solution space, we leave it for future investigation.}. Therefore, we have to set $B=O(r^{-2})$. When using the first equation of \eqref{Lorentz}, $B=0$. Consequently, all the sub-leadings of $A$, $E$, $u$, $r$, and $x^A$ are zero. This gives
\be\begin{split}
&r=e^{E_{R_0}(u',z',\bz')}r'+r_0(u',z',\bz'),\;\;\;\;u=u_0(u',z',\bz'),\;\;\;\;x^A=x^A_0(u',z',\bz'),\\
&A=A_0(u',z',\bz'),\;\;\;\;E=E_0(u',z',\bz'),\;\;\;\;B=0.
\end{split}\ee
To proceed, we check the asymptotic behavior of the transformed tetrad. From $n'$, we get
\be\begin{split}
&n'^{u'}=e^{-E_{R0}}\p_u u'_0+O(r^{-1}),\\
&n'^{r'}=e^{-2E_{R0}}\Lambda r^2+O(r),\\
&n'^{z'}=e^{-E_{R0}}\p_u z'_0+O(r^{-1}).\\
\end{split}\ee
This implies
\be
E_{R0}=0,\;\;\;\;u'_0=u+u'_0(z,\bz),\;\;\;\;z'_0=Y(z,\bz).
\ee
We continue to check $m'$,
\be\begin{split}
&m'^{u'}=e^{iE_I}(P\bp u'_0 + Q \p u'_0) + O(r^{-2}),\\
&m'^{r'}=A_0 + O(r^{-1}),\\
&m'^{z'}=O(r^{-1}).\\
\end{split}\ee
This leads to
\be
A_0=0,\;\;\;\;u'_0=u+c,\;\text{where $c$ is a real constant}.
\ee
The condition that the terms of $\frac{1}{r'^2}$ in $\rho'$ is absent yields $r_0=0$. Since we did not require the boundary 2 metric to be conformally flat, there is no more constraint on $Y$ and $\bY$. The full residual gauge transformation is a Diff($S^2$) and a third class of rotation $m'=e^{iE_{I_0}}m$ as well as a translation in the time direction. All the residual transformations are performed on the boundary 2 surface. The action on the boundary 2 surface is very simple:
\be
Q'=e^{iE_I}\eth Y,\;\;\;\;P'=e^{iE_I}\eth\bY.
\ee

\section{Flat limit and constant foliation}\label{flat}
A flat limit of the solution space in section \ref{solution} can be taken directly by setting $\Lambda=0$. However, from \eqref{sigma}, $P$ and $Q$ will have the following constraint
\be
Q\p_u P= P \p_u Q.
\ee
This leads to $Q=P \tilde Q$, and $\tilde Q$ is a function of $(z,\bz)$. Newman and Unti \cite{Newman:1962cia} have set the boundary 2 metric to be conformally flat, namely, $\tilde Q=0$. Removing the condition on the boundary 2 metric, the solution space of NP equations can be even larger\footnote{\pu{To be more precisely, the boundary 2 metric was set to be conformally flat before solving the non-radial equations in \cite{Newman:1962cia}. Hence the flat limit of the solution space in section \ref{solution} is larger than \cite{Newman:1962cia} in the sense that $\alpha^0$ and $\gamma^0$ are more general.}}. Via a Weyl transformation, one can reorganize the flat solution space in the present work, such that both $P$ and $Q$ are u-independent \pu{\textit{i.e.} choosing a constant foliation}. \pj{The residual gauge transformation of this case is supertranslation$\ltimes$Diff($S^2$) since the boundary 2 metric is not conformally flat. This is the phase space discussed in \cite{Campiglia:2014yka,Campiglia:2015yka,Compere:2018ylh} and it has important applications in the study of the equivalence between asymptotic symmetries and soft graviton theorems \cite{Kapec:2014opa}.}

\pu{With a cosmological constant, as we have shown in the previous section, the residual gauge transformation only consists of a Diff($S^2$) and a third class of rotation (somehow half of Weyl transformation)\footnote{\pu{Here, we only deal with transformations in the NP formalism. In the conformal frame \cite{Penrose:1962ij}, there is considerable freedom in the choice of the conformal factor. It is of interest to study the effect of the residual conformal freedom in NP formalism elsewhere.}}. Accordingly, it is not possible to reorganize the solution space to have both $P$ and $Q$ u-independent (hence $\gamma^0=0$) by residual gauge transformation. One can only set Im$(\gamma^0)=0$ via a third class of rotation. This is another reason to remove the constant foliation condition $\gamma^0=0$ in \cite{Saw:2016isu,Saw:2017hsf}. Though any boundary 2 metric is connected by Diff($S^2$), most elements of the Diff($S^2$) transformations are singular and will arise (new) topological degree of freedom. If we just focus on the issues of gravitational radiation in (A)dS space-times, it is better to restrict ourselves to the Lorentz transformations.}

\section{Discussions}
In this work, we have derived the solution space of the Newman-Penrose formalism with a cosmological constant. The residual gauge transformation that preserves the solution space has also been worked out. The solution space has a well-defined flat limit. The residual gauge transformation of the resulting solution space consists of supertranslation$\ltimes$Diff($S^2$) and this phase space should have certain relevance to soft graviton theorems according to recently discovered equivalence between soft theorems and asymptotic symmetries \cite{He:2014laa}.

There are several interesting questions about the solution space with a cosmological constant that need to be addressed in the future:
\begin{itemize}
\item
The characteristic initial value problem: In the case $\Lambda=0$, a solution of NP equations is determined by spesifying the news function $\sigma^0$ and conformal factor $P$ at any time $u$; $\Psi_1^0$, $\Psi_2^0+\xbar\Psi_2^0$, and $\Psi_0$ (the entire series) at the initial time $u_0$. According to \eqref{evolution}, the time evolution equations of Weyl tensors are mixed due to the appearance of the cosmological constant. Hence, the initial data that determine a solution are not yet fully understood.
\item
Newman-Penrose conserved quantities: In \cite{Newman:1968uj}, an infinite number of gravitationally-conserved quantities was discovered from the NU solution space. Once the characteristic initial value problem is solved in the case with a cosmological constant, gravitationally-conserved quantities should be constructed as well.
\item
Bondi mass and mass-loss: The analogue of the Bondi mass aspect and the mass-loss formula with respect to the solution space need to be stressed elsewhere. A second relevant (generalized) problem is to study the full current algebra of the asymptotic symmetries group \cite{tocome}.
\item
Polyhomogeneous series: Expansion with logarithms can be applied to derive a larger solution space. Correspondingly, one should apply the asymptotic behavior $\Psi_0=\frac{\Psi_0^0}{r^5}+o(r^{-5})$ rather than $\Psi_0=\frac{\Psi_0^0}{r^5}+O(r^{-6})$ in the present treatment.
\end{itemize}

\section*{Acknowledgments}

The author thanks Glenn Barnich, Vee-Liem Saw, and Xiaoning Wu for useful discussions and additionally Glenn Barnich again for suggesting this problem.  Special thanks are due to the anonymous referee for valuable suggestions which are very helpful in improving the original manuscript. This work is supported in part by the China Postdoctoral Science Foundation (Grant No. 2017M620908), by the National Natural Science Foundation of China (Grant No. 11575202).

\appendix

\section{Useful relations in Newman-Penrose formalism}
\label{relation}
From the orthogonality conditions and normalization conditions of the basis vectors, one can obtain the following relations
\be
\begin{split}\label{1}
&l^\nu \n_\nu l_\mu=(\epsilon+\bar \epsilon)l_\mu -\kappa \bm_\mu - \bar \kappa m_\mu,\\
&n^\nu \n_\nu l_\mu=(\gamma+\bar \gamma)l_\mu -\tau \bm_\mu - \bar \tau m_\mu,\\
&m^\nu \n_\nu l_\mu=(\beta+\bar \alpha)l_\mu -\sigma \bm_\mu - \bar \rho m_\mu,\\
&\bm^\nu \n_\nu l_\mu=(\alpha + \bar \beta)l_\mu - \rho \bm_\mu -\bar\sigma m_\mu,
\end{split}
\ee
\be
\begin{split}
&l^\nu \n_\nu n_\mu=-(\epsilon+\bar \epsilon)n_\mu + \bar\pi \bm_\mu + \pi m_\mu,\\
&n^\nu \n_\nu n_\mu=-(\gamma+\bar \gamma)n_\mu + \bar\nu \bm_\mu + \nu m_\mu,\\
&m^\nu \n_\nu n_\mu=-(\beta+\bar \alpha)n_\mu + \bar\lambda \bm_\mu + \mu m_\mu,\\
&\bm^\nu \n_\nu n_\mu=-(\alpha + \bar \beta)n_\mu + \bar\mu \bm_\mu + \lambda m_\mu,\\
\end{split}
\ee
\be
\begin{split}
&l^\nu \n_\nu m_\mu=(\epsilon - \bar\epsilon) m_\mu - \kappa n_\mu + \bar\pi l_\mu,\\
&n^\nu \n_\nu m_\mu=(\gamma - \bar\gamma) m_\mu  - \tau n_\mu + \bar\nu l_\mu,\\
&m^\nu \n_\nu m_\mu= (\beta - \bar\alpha) m_\mu - \sigma n_\mu + \bar\lambda l_\mu,\\
&\bm^\nu \n_\nu m_\mu= (\alpha - \bar\beta) m_\mu - \rho n_\mu + \bar\mu l_\mu,
\end{split}
\ee

\section{NP equations}
\label{NP}
\textbf{Radial equations}
\bea
&&D\rho =\rho^2+\sigma\xbar\sigma,\label{R1}\\
&&D\sigma=2\rho \sigma + \Psi_{0},\label{R2}\\
&&D\tau =\tau \rho +  \xbar \tau \sigma   + \Psi_1 ,\label{R3}\\
&&D\alpha=\rho  \alpha + \beta \xbar \sigma  ,\label{R4}\\
&&D\beta  =\alpha \sigma + \rho  \beta + \Psi_{1},\label{R5}\\
&&D\gamma=\tau \alpha +  \xbar \tau \beta  + \Psi_2 - \Lambda,\label{R6}\\
&&D\lambda=\rho\lambda + \xbar\sigma\mu ,\label{R7}\\
&&D\mu =\rho \mu + \sigma\lambda + \Psi_{2} + 2\Lambda,\label{R8}\\
&&D\nu =\xbar\tau \mu + \tau  \lambda + \Psi_3,\label{R9}\\
&&DU=\xbar\tau\omega+\tau\xbar\omega - (\gamma+\xbar\gamma),\label{R10}\\
&&DX^A=\xbar\tau L^A + \tau\bar L^A,\label{R11}\\
&&D\omega=\rho\omega+\sigma\xbar\omega-\tau,\label{R12}\\
&&DL^A=\rho L^A + \sigma \bar L^A,\label{R13}\\
&&D\Psi_1 - \xbar\delta \Psi_0 =  4 \rho \Psi_1 - 4\alpha \Psi_0,\label{R14}\\
&&D\Psi_2 - \xbar\delta \Psi_1 =   3\rho \Psi_2  - 2 \alpha \Psi_1- \lambda \Psi_0,\label{R15}\\
&&D\Psi_3 - \xbar\delta \Psi_2 =  2\rho \Psi_3 - 2\lambda \Psi_1,\label{R16}\\
&&D\Psi_4 - \xbar\delta \Psi_3 = \rho  \Psi_4 + 2 \alpha \Psi_3 - 3 \lambda \Psi_2.\label{R17}
\eea

\textbf{Non-radial  equations}
\bea
&&\Delta\lambda  = \xbar\delta\nu- (\mu + \xbar\mu)\lambda - (3\gamma - \xbar\gamma)\lambda + 2\alpha \nu - \Psi_4,\label{H1}\\
&&\Delta\rho= \xbar\delta\tau- \rho\xbar\mu - \sigma\lambda  -2\alpha \tau + (\gamma + \xbar\gamma)\rho  - \Psi_2 -2\Lambda,\label{H2}\\
&&\Delta\alpha = \xbar\delta\gamma +\rho \nu - (\tau + \beta)\lambda + (\xbar\gamma - \gamma -\xbar \mu)\alpha  -\Psi_3 ,\label{H3}\\
&&\Delta \mu=\delta\nu-\mu^2 - \lambda\xbar\lambda - (\gamma + \xbar\gamma)\mu   + 2 \beta \nu ,\label{H4}\\
&&\Delta \beta=\delta\gamma - \mu\tau + \sigma\nu + \beta(\gamma - \xbar\gamma -\mu) - \alpha\xbar\lambda ,\label{H5}\\
&&\Delta \sigma=\delta\tau - \sigma\mu - \rho\xbar\lambda - 2 \beta \tau + (3\gamma - \xbar\gamma)\sigma  ,\label{H6}\\
&&\Delta \omega=\delta U +\xbar\nu -\xbar\lambda\xbar\omega + (\gamma-\xbar\gamma-\mu)\omega,\label{H7}\\
&&\Delta L^A=\delta X^A - \xbar\lambda \bar L^A + (\gamma-\xbar\gamma-\mu)L^A,\label{H8}\\
&&\delta\rho - \xbar\delta\sigma=\rho\tau - \sigma (3\alpha - \xbar\beta)   - \Psi_1 ,\label{H9}\\
&&\delta\alpha - \xbar\delta\beta=\mu\rho - \lambda\sigma + \alpha\xbar\alpha + \beta\xbar\beta - 2 \alpha\beta - \Psi_2  + \Lambda,\label{H10}\\
&&\delta\lambda - \xbar\delta\mu= \mu \xbar\tau + \lambda (\xbar\alpha - 3\beta) - \Psi_3 ,\label{H11}\\
&&\delta \xbar\omega-\bar\delta\omega=\mu - \xbar\mu - (\alpha - \xbar\beta) \omega +  (\xbar\alpha - \beta)\xbar\omega,\label{H12}\\
&&\delta \bar L^A - \bar\delta L^A= (\xbar\alpha - \beta)\bar L^A -  (\alpha - \xbar\beta) L^A ,\label{H13}\\
&&\Delta\Psi_0 - \delta \Psi_1 = (4\gamma -\mu)\Psi_0 - (4\tau + 2\beta)\Psi_1 + 3\sigma \Psi_2,\label{H14}\\
&&\Delta\Psi_1 - \delta \Psi_2 = \nu\Psi_0 + (2\gamma - 2\mu)\Psi_1 - 3\tau \Psi_2 + 2\sigma \Psi_3 ,\label{H15}\\
&&\Delta\Psi_2 - \delta \Psi_3 = 2\nu \Psi_1 - 3\mu \Psi_2 + (2\beta - 2\tau) \Psi_3 + \sigma \Psi_4,\label{H16}\\
&&\Delta\Psi_3 - \delta \Psi_4 = 3\nu \Psi_2 - (2\gamma + 4\mu) \Psi_3 + (4\beta - \tau) \Psi_4.\label{H17}
\eea

\section{\pu{Details in solving NP equations}}
\label{process}
\pu{
The method of solving NP equations was originally schemed in \cite{Newman:1961qr}, later implemented by Newman and Unti in detail \cite{Newman:1962cia}. We follow exactly the derivation of Newman and Unti. The radial equations are solved in different groups. The first group is \eqref{R1} and \eqref{R2}. Those equations are not affected by $\Lambda$. Once the whole series of $\Psi_0$ is given as initial data by \eqref{psi0}, $\rho$ and $\sigma$ are solved out as \eqref{rho} and \eqref{sigma}. Inserting the solutions of $\rho$ and $\sigma$ into \eqref{R13}, one gets $L^A$ as \eqref{Lz} and \eqref{Lzb}, then $L_A$ by the condition $L_AL^A=0,\;L_A\bar L^A=-1$. The second group of radial equations consists of \eqref{R4}, \eqref{R5}, \eqref{R12} and \eqref{R14}. Those equations are not modified by $\Lambda$ either. One can work out $\alpha$, $\beta$, $\omega$ and $\Psi_1$ as \eqref{alpha}, \eqref{beta}, \eqref{omega}, and \eqref{psi1} respectively, then $\tau$ from gauge condition $\tau=\xbar\alpha+\beta$. Inserting $\tau$ and $L^A$ into \eqref{R11}, $X^A$ is obtained as \eqref{XA}. Until now, the cosmological constant has not appeared yet. We are just repeating the result of \cite{Newman:1962cia}. We continue with the third group of radial equations that include \eqref{R7}, \eqref{R8}, and \eqref{R15}. One can see the cosmological constant appears for the first time in \eqref{R8}. But one can just apply the same method as the first two groups to solve out $\mu$, $\lambda$, and $\Psi_2$, which are \eqref{mu}, \eqref{lambda}, and \eqref{psi2}. Then, $\gamma$ is derived from \eqref{R6} as \eqref{gamma},  $U$ is derived from \eqref{R10} as \eqref{U},  $\Psi_3$ is derived from \eqref{R16} as \eqref{psi3},  $\nu$ is derived from \eqref{R9} as \eqref{nu}, and finally $\Psi_4$ is derived from \eqref{R17} as \eqref{psi4}.
}

\pu{
The solutions to the radial equations are less affected by cosmological constant. Now we will check the constraints from non-radial equations on the integration constants of the radial equations. By inserting the solutions of the radial equations into the non-radial equations, the leading term of the Bianchi identities \eqref{H14}-\eqref{H17} will lead to the time evolution equations \eqref{evolution}. The Weyl tensors are more entangled because of $\Lambda$. More constraints are obtained from the leading term of the rest of the non-radial equations:
}

\pu{
\eqref{H13} yields \eqref{alpha0}.
}

\pu{
\eqref{H9} yields \eqref{omega0}
}

\pu{
\eqref{H11} yields \eqref{Psi30}.
}

\pu{
\eqref{H10} yields \eqref{mu0}.
}

\pu{
\eqref{H12} yields the magnetic part of $\Psi_2^0$ as \eqref{Psi20}.
}

\pu{
\eqref{H8} yields \eqref{sigma0} and \eqref{gamma0}.
}

\pu{
\eqref{H2} yields \eqref{U0}.
}

\pu{
\eqref{H6} yields \eqref{lambda0}.
}

\pu{
\eqref{H1} yields \eqref{Psi40}.
}

\pu{
\eqref{H7} yields \eqref{nu0}.
}

\pu{
\eqref{H4} and \eqref{H3} yield two identities \eqref{identities}.
}

\pu{The last one \eqref{H5} just yields the complex conjugate of the second identity in \eqref{identities}. Though it is not completely clear that if there is new information coming from higher powers of $\frac1r$\footnote{\pu{There is no new information coming from higher powers of $\frac1r$ in the $\Lambda=0$ case \cite{Newman:1962cia}.}}, we computed the next-to-leading order of \eqref{H2}, \eqref{H6}, \eqref{H8}-\eqref{H10}, \eqref{H12}, \eqref{H13} and found no more information. Due to the tediousness of the computation, we did not continue with other equations and higher orders. Some arguments from the structure of NP equations should be made to justify that all higher orders have no more information.}

\bibliography{ref}

\bibliographystyle{utphys}

\end{document}